# Ultrafast Vibrational Control of Hybrid Perovskite Devices Reveals the Influence of the Organic Cation on Electronic Dynamics

Nathaniel. P. Gallop[†], Dmitry R. Maslennikov[†], Katelyn P. Goetz[‡], Zhenbang Dai[Δ], Aaron M. Schankler[Δ], Woongmo Sung[°], Satoshi Nihonyanagi[°], Tahei Tahara[°], Maryna Bodnarchuk[§], Maksym Kovalenko[§], Yana Vaynzof[‡¥], Andrew M. Rappe[Δ], Artem A. Bakulin[†*]

† Department of Chemistry, Imperial College London, 83 Wood Ln, London, United Kingdom W12 0BZ

‡ Chair for Emerging Electronic Technologies, Technical University of Dresden, Nöthnitzer Str. 61, 01187 Dresden, Germany

¥ Leibniz Institute for Solid State and Materials Research Dresden, Helmholtzstraße 20, 01069 Dresden, Germany

Δ Department of Chemistry, University of Pennsylvania, Philadelphia, Pennsylvania 19104-6323, USA

§ Dept. of Chemistry and Applied Biosciences, ETH Zurich, 8093 Zurich, Switzerland

° Molecular Spectroscopy Laboratory, RIKEN, 2-1 Hirosawa, Wako, Saitama 351-0198, Japan

*a.bakulin@imperial.ac.uk

Vibrational control (VC) of photochemistry through the optical stimulation of structural dynamics is a nascent concept only recently demonstrated for model molecules in solution. Extending VC to state-of-the-art materials may lead to new applications and improved performance for optoelectronic devices. Metal halide perovskites are promising targets for VC due to their mechanical softness and the rich array of vibrational motions of both their inorganic and organic sublattices. Here, we demonstrate the ultrafast VC of $FAPbBr_3$ perovskite solar cells via intramolecular vibrations of the formamidinium cation using spectroscopic techniques based on vibrationally promoted electronic resonance. The observed short (~300 fs) time window of VC highlights the fast dynamics of coupling between the cation and inorganic sublattice. First-principles modelling reveals that this coupling is mediated by hydrogen bonds that modulate both lead halide lattice and electronic states. Cation dynamics modulating this coupling may suppress non-radiative recombination in perovskites, leading to photovoltaics with reduced voltage losses.

## Main

Highly selective, on-demand photocontrol of the electronic dynamics and reactivity of a material is one of the holy grails of photochemistry. One promising avenue to realize this control is through the selective stimulation of a material's structural and vibrational dynamics[1]. However, despite considerable effort over the last two decades, achieving vibrational control (VC) of photoreactivity has proven to be a challenging task, requiring advances in chemistry, materials science and ultrafast photonics. Only a handful of studies to date have demonstrated the utility of VC; examples include the mode-selective modulation of charge transfer and carrier detrapping[2,3,4,5], triggered chemical reactions and species-selective photochemistry[6,7]. However, in these pioneering experiments, VC has largely been limited to model chemical systems in solution; the application of VC to practically useful material systems has only been demonstrated on trapped carriers with approximately millisecond lifetimes, disregarding the ultrafast character and selectivity of the approach[3]. Ultrafast VC measurements would provide invaluable insights into important photophysical phenomena, such as electron–phonon-coupling effects, polaronic effects and ultrafast structural dynamics, which, in turn, may lead to improvements in material properties and device performances.

One of the most promising classes of optoelectronic materials are 'hybrid' organic–inorganic perovskites (HOIPs)—high-performance photovoltaic materials that are inexpensive to produce, lightweight, flexible and broadly tunable[8]. HOIP nanomaterials also demonstrate potential in diverse applications including light-emitting diodes[9], lasers[10] and non-volatile memory[11]. Moreover, as a hybrid material, HOIPs combine the low electronic disorder and delocalized electronic states of inorganic semiconductors with the localized and tunable structural dynamics of organic molecules. The combination of organic and inorganic sublattices results in a rich vibrational manifold, from low-energy phonons to high-frequency intramolecular vibrations. These features, together with the rotational mobility of the organic cation[12,13,14,15] and high structural anharmonicity[16,17], result in a complex dynamical behaviour well suited for VC experiments. Such experiments can, for instance, address interactions between the organic A-site cations and the surrounding inorganic cage. Despite minimally contributing to the HOIPs' frontier electronic states[18], organic cation dynamics may influence the optoelectronic properties of HOIPs through a variety of different means, including via the modulation of dielectric constant[19] and large polaron formation[20,21], among others[22,23]. Several experimental studies have suggested that interactions between the organic A-site cation and the inorganic cage may modulate the band-edge properties of the perovskite[24,25,26,27]. However, the actual mechanisms behind these phenomena remains controversial, and the extent of their influence on the optoelectronic properties of HOIPs remains unclear[28,29,30,31].

Here we address the issue of cation–lattice interactions in HOIPs via the VC of perovskite photovoltaic devices. For this, we realize a novel spectroscopic technique based on the vibrationally promoted electronic resonance (VIPER) approach. Conceived as an optical extension to two-dimensional (2D) exchange spectroscopy experiments[6,7], the VIPER approach exhibits several advantages that make it an ideal candidate for exploiting VC, including high infrared (IR) selectivity, background-free detection, sub-picosecond time resolution and sensitivity to electron–phonon couplings[6,7]. To more directly access material and device properties, we have combined VIPER with photocurrent detection (termed PC-VIPER) (Fig. 1a). In this approach, interferometrically separated IR pulses first arrive at the photovoltaic device, bringing resonant IR-active vibrational modes into their excited states. After a delay time (*t*), a visible pulse promotes this sub-ensemble to the electronic first excited state, which can then produce a photocurrent. The visible pulse chosen here is off-resonant with this electronic transition such that the visible pulse alone does not electronically excite the sample; only the combination of vibrational and electronic transitions populates the electronic excited state and produces photocurrent. Moreover, this process can only occur if the target vibrational mode is coupled to the electronic transition in question and if this coupling persists during the time interval between the vibrational and electronic transitions. This makes any observed signal intrinsically sensitive to electron–phonon-coupling effects[6,7]. Furthermore, changing the time delay between the visible and IR pulses enables the timescales of these couplings and/or the lifetimes of the stimulated vibrational modes to be monitored. Additionally, the use of an interferometer enables resolution along the IR frequency axis; by sweeping the interferometer and Fourier transforming the resulting time-domain data, any signal can be assigned to individual vibrational bands within the bandwidth of the IR pulse. Our approach also benefits from all the advantages of photocurrent action spectroscopy[32], including operando measurements at low-power conditions, the absence of scattering and outcoupling artefacts, as well as the use of sensitive background-free lock-in detection. Although PC-VIPER is suited for perovskite materials used for photovoltaics; we also demonstrate that luminescent perovskite nanocrystals (NCs) for light-emitting diode and fluorescence-labelling applications can be studied by an analogous technique in which material photoluminescence is detected after the VIPER excitation sequence (termed PL-VIPER). PL-VIPER is broadly analogous to the recently reported fluorescence-encoded infrared technique that has demonstrated high sensitivity (down to single molecules) for model molecular systems[33,34].

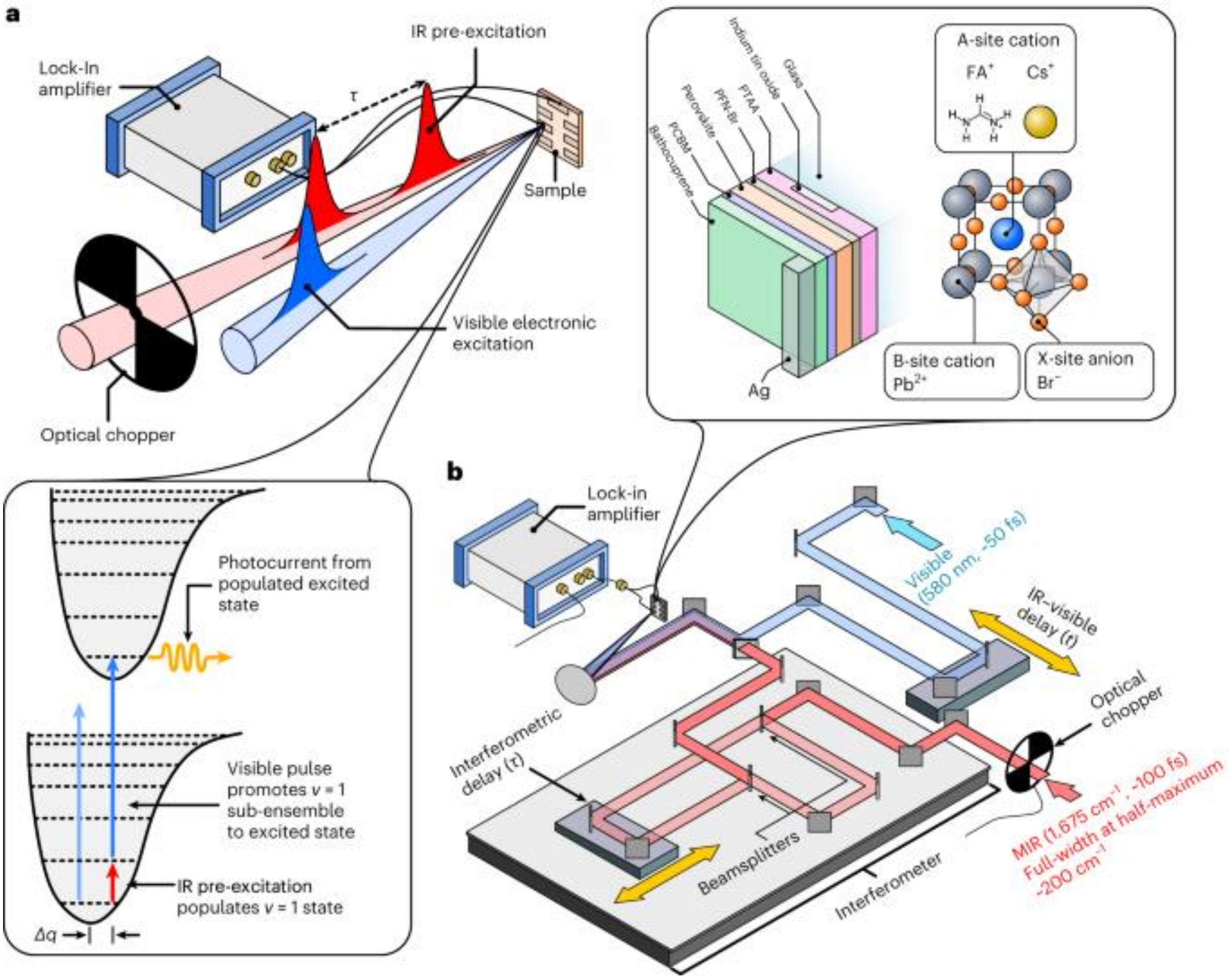

**Figure 1. a**, Simplified experimental setup and pulse sequence for the PC-VIPER experiment (the inset shows the state diagram illustrating the PC-VIPER double-resonance response, with $\Delta q$ representing the change in nuclear coordinates upon excitation). **b**, Practical implementation of the PC-VIPER experiment, incorporating an interferometer for IR pre-excitation (the inset shows the architecture and structure of the perovskite device employed in these experiments).

The systems under investigation here include bulk $FAPbBr_3$ and $CsPbBr_3$ perovskite devices as well as $FAPbBr_3$ nanocrystalline films. The incident photon to converted electron spectra of the $FAPbBr_3$ device (Fig. 2a) exhibits a sharp band edge at 550 nm with efficiencies up to 80%. To ensure minimal direct absorption of the visible pump pulse without IR pre-excitation, the centre wavelength of the pump was set at approximately 580 nm, well below the onset of efficient photocurrent generation.

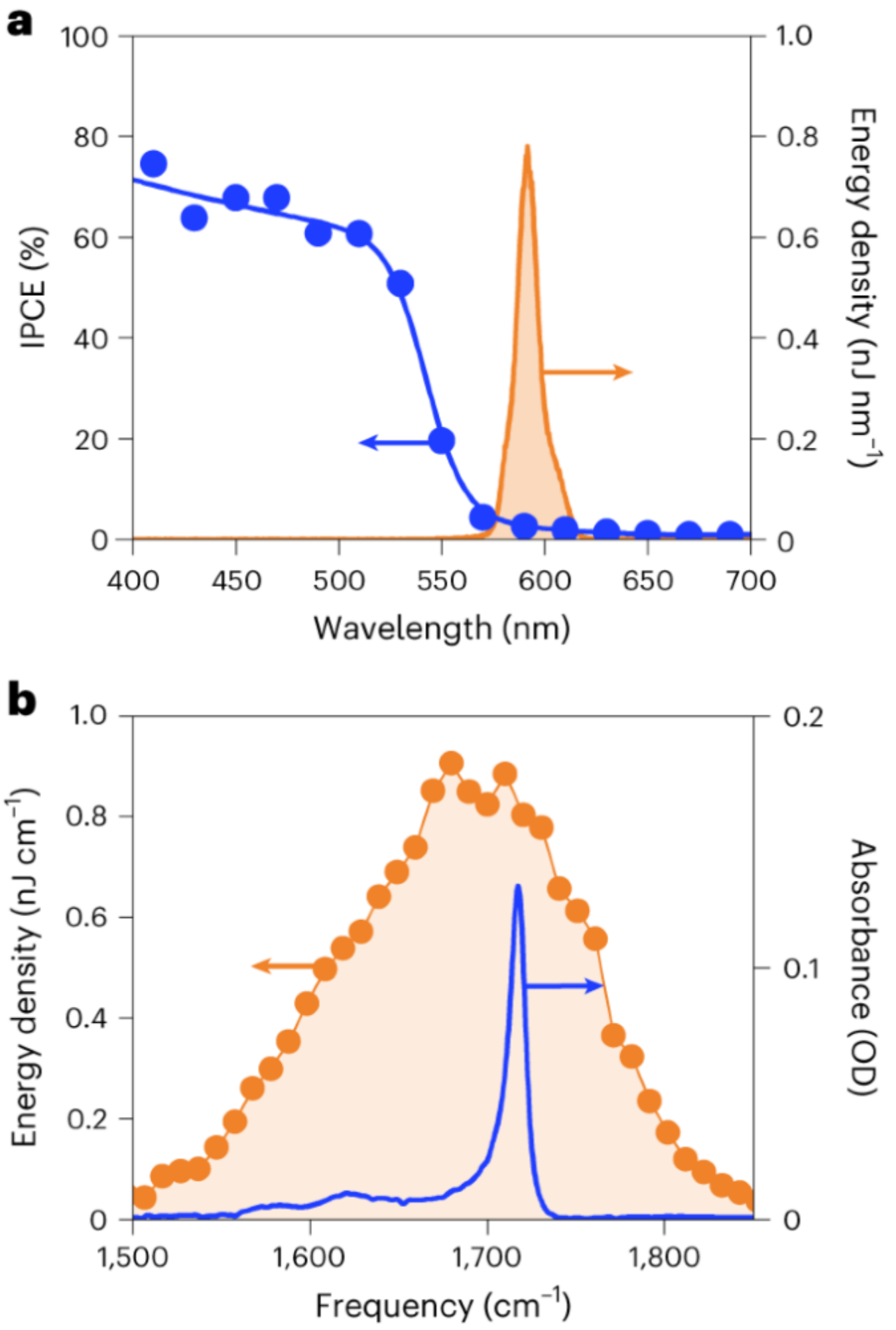

**Figure 2. a**, Incident photon to converted electron (IPCE) spectrum of an $FAPbBr_3$ device (blue), along with the spectrum of the visible electronic excitation pulse (red shaded) employed in this work. **b**, IR absorption spectrum of $FAPbBr_3$ (blue), overlaid on the spectrum of the IR pre-excitation pulse (red) employed in this experiment.

The FTIR spectrum of an $FAPbBr_3$ film (Fig. 2b) shows a weak band at ~1,615 cm$^{-1}$ and a strong line at ~1,715 cm$^{-1}$, corresponding to N–H scissoring (δ1($NH_2$)) and asymmetric stretching of the N–C=N double bond (ν(N–C=N)) of formamidinium (FA), respectively[35]. The IR attenuated total reflection spectra of the full $FAPbBr_3$ and $CsPbBr_3$ devices (Supplementary Note 1) show no other vibrational modes over this measurement range. To avoid IR absorption and dispersion in glass, indium tin oxide and poly[bis(4-phenyl)(2,4,6-trimethylphenyl)amine] (PTAA) charge extraction layer, we illuminate the perovskite through semitransparent top Ag electrodes during VIPER experiments.

**PC-VIPER and PL-VIPER experiments**

Figure 3 displays the 2D PC-VIPER maps for the $FAPbBr_3$ and $CsPbBr_3$ photovoltaic devices. The maps are obtained by scanning the interferometer delay (τ) for a range of IR–visible delays (t). Thereafter, the time-domain signal is Fourier transformed along the τ axis to produce an IR action spectrum[3,36].

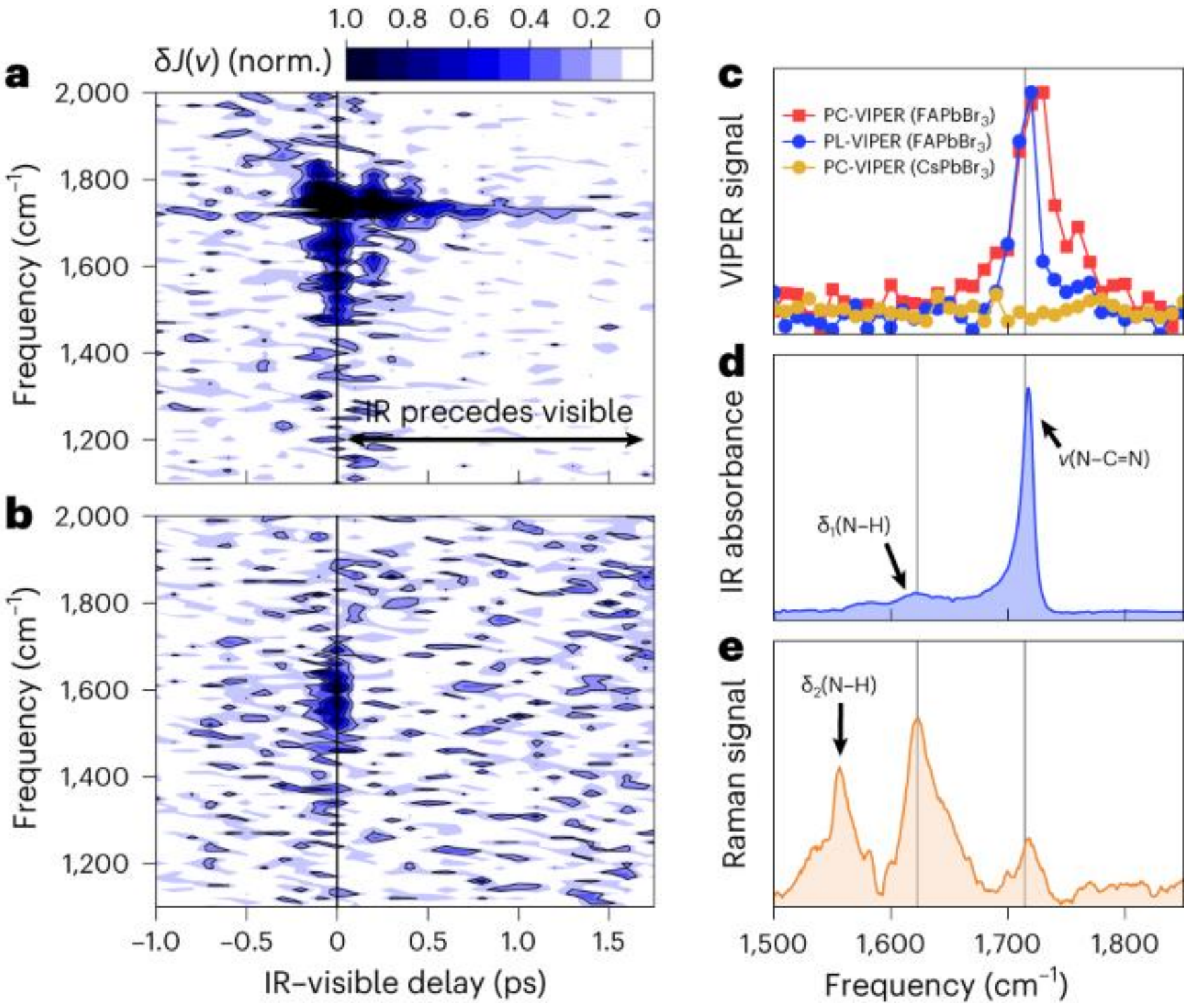

**Figure 3. a**,**b**, PC-VIPER maps for $FAPbBr_3$ (**a**) and $CsPbBr_3$ (**b**). The black arrow in **a** denotes where the IR pulse precedes the visible pulse, whereas the horizontal dashed lines denote the approximate spectral width (>10% peak amplitude) of the IR pulse. **c**–**e**, Assignment of the PC-VIPER data to known vibrational modes: one-dimensional PC-VIPER spectra for $FAPbBr_3$and $CsPbBr_3$ (**c**); IR spectrum of $FAPbBr_3$ (**d**); Raman spectrum of $FAPbBr_3$ (**e**). The frequencies of $\delta_1(NH_2)$ and *v*(N–C=N) modes (approximately 1,620 and 1,715 $cm^{-1}$, respectively) are demarcated in grey. A one-dimensional PL-VIPER spectrum (Fig. 4a) is also displayed in **c**; in all these cases, the one-dimensional traces are obtained by averaging the 2D maps in Figs. 3a,b and 4a between 100 and 1,000 fs.

Within the pulse-overlap regime (±50 fs), both materials exhibit a broad, featureless signal centred at ~1,600 $cm^{-1}$, attributable to residual non-resonant two-photon effects. This broad signal was greatly suppressed via the time-domain phasing and filtering process (Supplementary Note 2) but could not be fully eliminated. Beyond the pulse-overlap time region (that is, >80 fs, where the IR vibrational pre-excitation arrives before the visible electronic pump), the non-resonant signal for $FAPbBr_3$(but crucially not $CsPbBr_3$) gives way to a single sharp line, centred at 1,720 (±10) $cm^{-1}$. The agreement of this line with the frequency of the *v*(N–C=N) vibration15 of FA (Fig. 2b), alongside the absence of a similar line for $CsPbBr_3$, suggests that it originates from FA vibrations. We assign this feature to the vibronic coupling between the FA cation's $\nu$(N–C=N) vibrational mode and the electronic state of $FAPbBr_3$. Interestingly, despite the IR pre-excitation spectrum covering two features (the $\nu$(N–C=N) and $\delta_1(NH_2)$ modes), only the line at 1,720 cm$^{-1}$ is observed. Intuitively, any VIPER-active modes should be both IR and Raman active, as IR activity enables the population of a vibrationally excited state and (resonance) Raman activity is associated with the strength of vibronic coupling. However, considering that we use non-resonance Raman activity for our analysis and given the complexity of the interaction mechanism, more complex selection rules may be in place.

To confirm that the observed vibronic phenomenon exists across different perovskite materials, we repeat our VC experiments on $FAPbBr_3$ NCs. Such NCs are known to have outstanding photoluminescent properties; for this reason, we employed PL-VIPER experiments (Supplementary Note 3). The experimental parameters and data analysis approach were identical to the PC-VIPER experiments described above.

As with PC-VIPER, a single sharp vibrational line at ~1,720 cm$^{-1}$, corresponding to the $\nu$(N–C=N) mode of FA, is observed in the PL-VIPER experiments (Fig. 4) with a distinct absence of the $\delta_1(NH_2)$ mode also covered by the pulse. The PC-VIPER and PL-VIPER responses exhibit similar lineshapes (Fig. 3c) and dynamics (Fig. 4b). The slight discrepancy in decay times (240 ± 30 fs from PL-VIPER versus 280 ± 20 fs for PC-VIPER) may be due to differing material morphologies and minor ligand effects in our $FAPbBr_3$ NCs compared with bulk $FAPbBr_3$. Nevertheless, these findings indicate that we monitor the same VC effect using both photocurrent and photoluminescence.

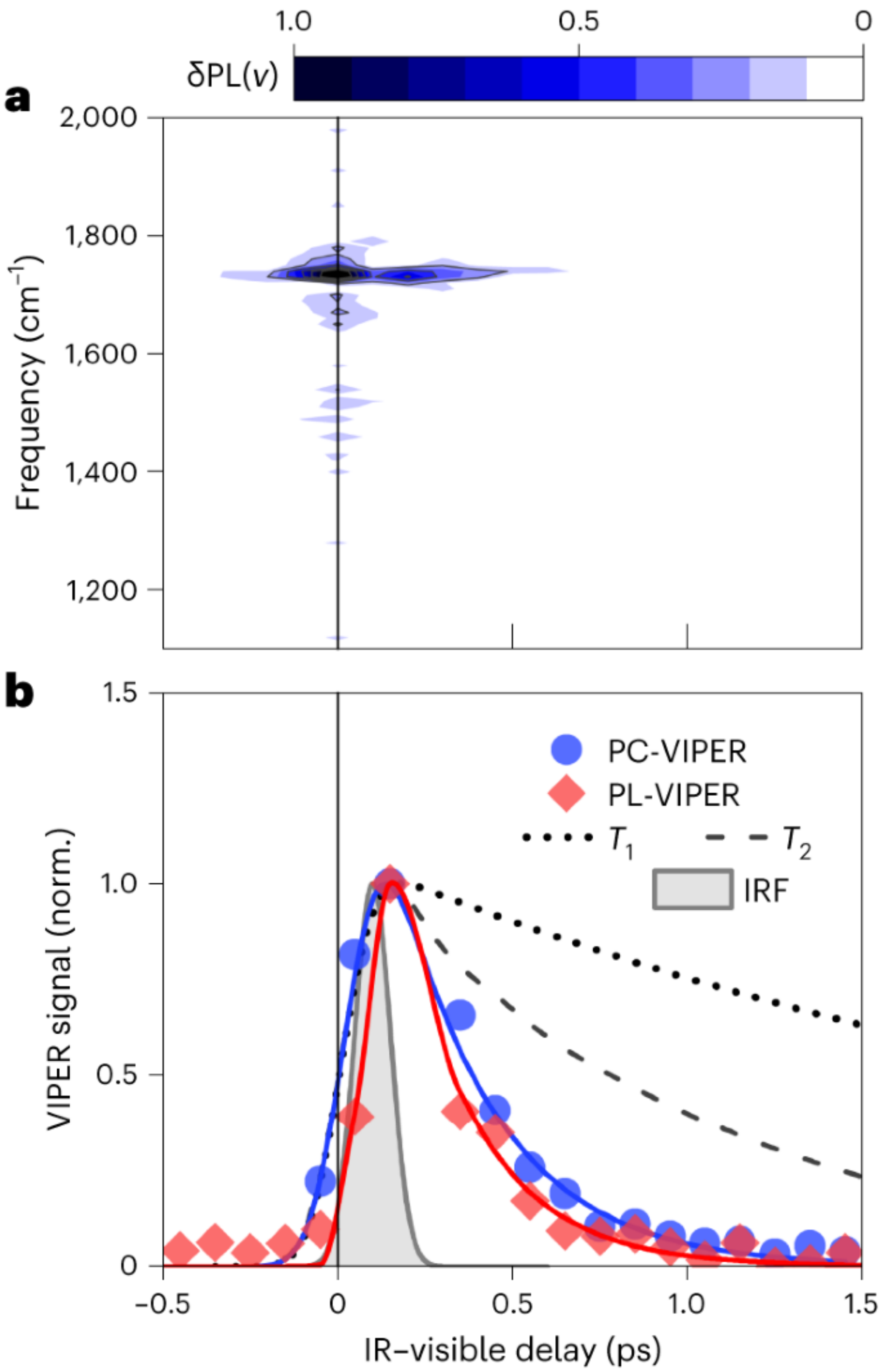

**Figure 4. a**, PL-VIPER map for $FAPbBr_3$. **b**, Kinetics of the PL-VIPER and PC-VIPER signals at 1,720 cm$^{-1}$; the population relaxation ($T_1$) and dephasing ($T_2$) times are shown for comparison. The instrument response function (IRF) in **b** was estimated from a Gaussian fit of the non-resonant response of $CsPbBr_3$.

We now address alternative explanations for the emergence of this signal. We compared the IR spectra of the full $FAPbBr_3$ and $CsPbBr_3$ devices (Supplementary Note 1) with those of the $FAPbBr_3$ films shown in Fig. 2 and found no substantial difference, indicating that the signal purely originates from the perovskite active layer. We can rule out an anti-Stokes Raman-like effect, in which the off-resonant visible photons are inelastically scattered from the excited vibrational state because the lifetimes of our signals are substantially shorter than the lifetime of the vibrational mode previously observed via 2D-IR spectroscopy15 or estimated from lineshape analysis (Fig. 4b). Similarly, the persistence of the VIPER signal is also too short to emerge from a vibrational sum-frequency generation (VSFG) effect, whose temporal profile is determined by the free induction decay of the vibration37,38,39. We estimate the free induction decay of the 1,720 $cm^{-1}$ mode to be ~850–870 fs, based on a lineshape analysis of the IR absorption spectrum (Supplementary Note 4), meaning that the corresponding estimated lifetime of the VSFG signal is approximately ~430 fs (half the free induction decay). This value is substantially longer than the 280 fs lifetime of our PC-VIPER and PL-VIPER signals (Supplementary Fig. 3c). Moreover, independently measured VSFG spectra of our $FAPbBr_3$ films exhibit a negligible signal at 1,720 $cm^{-1}$ (Supplementary Note 5), and VSFG signals were also not observed in wavelength-dependent PL-VIPER control experiments (Supplementary Note 6). Collectively, these results suggest that the VIPER signal decay is controlled by some other process, rather than simply the population or dephasing of the vibrational state. Finally, sample-heating[31] effects on the band edge can be discarded due to the rapidity with which the observed signals grow in (<100 fs) and dissipate (sub-picosecond; >1 ns for sample-heating effects). Consequently, we conclude that vibronic coupling between the cation and electronic state of the perovskite provides the most plausible explanation for the observed VC phenomenon.

**The origin of vibronic coupling**

The VIPER kinetics reveal the nature of the vibronic-coupling phenomenon behind the PC-VIPER and PL-VIPER signals. The decays of both the PC-VIPER and PL-VIPER signals (Fig. 4b) are almost ten times faster than the population relaxation rate for the same mode previously observed for $FAPbI_3$ (2.8 ps) (ref. [15]) and roughly three times faster than the vibrational free induction decay (Supplementary Fig. 3b), suggesting that its dynamics depend on other processes within the material. In particular, we note that the relaxation rates of our VIPER signals closely match the timescales of the rotational dynamics of the FA cation. Cation rotation, given the highly directional nature of FA–Br hydrogen bonds[40], can strongly influence the interaction between the FA ions and the inorganic lattice, which determines the frontier electronic states of the perovskite. Intriguingly, the ~280 fs lifetimes of VIPER signals are comparable with the 300–400 fs ‘wobbling times’ of the rotational motions of the organic cartions observed for methylammonium and FA perovskites[15]. This indicates that the reorientation of the organic cation may dictate the timescales over which the vibrational states of the cation and electronic states of the perovskite lattice are coupled. This is supported by previous studies highlighting the importance of hydrogen bonding on the structural dynamics of the $PbBr_3$ sublattice[41], as well as a recent 2D-IR report that finds agreement between the rate of spectral diffusion of methylammonium-containing perovskites and the faster 'wobbling' timescales of the cations[42].

Identifying the mechanism of the vibronic coupling evolution should take into account the electronic structure of the perovskite, the rotational and vibrational dynamics of the FA cations and the interaction between the cations and inorganic sublattice. We anticipate that the kinetics of the VIPER signal depend on the dynamics of electron–phonon coupling, which can be affected by all of the above processes. To gain an atomistic insight into the origin of the vibronic-coupling phenomena observed in our PC-VIPER and PL-VIPER experiments, we performed a series of ab initio molecular dynamics (AIMD) simulations on a $2 \times 2 \times 2$ perovskite supercell (Supplementary Note 7).

We first used molecular dynamics simulations to quantify the hydrogen bond dynamics and correlate them with the dynamics of vibronic coupling observed in VIPER experiments. At each time step of the simulation, we calculated the distances between the four hydrogen atoms on the $–NH_2$ groups of FA and their nearest-neighbour $Br^-$ ion. The dynamics of the H•••Br bond distance predominantly consist of fluctuations about a local equilibrium value, with occasional drops in the H–Br distance that are almost immediately followed by a return to equilibrium (Supplementary Figs. 11 and 12). Previous ab initio studies have ascribed this behaviour to the stochastic formation and breakage of hydrogen bonds between the FA and surrounding $PbBr_3$ cage41. A statistical analysis of these events yields the ensemble-averaged dynamics of hydrogen bond breakage, expressed as the evolution of the N–H•••Br separation following hydrogen bond formation (Supplementary Fig. 8). A rapid initial increase in the bond distance shortly after hydrogen bond formation accounts for much of the H•••Br bond evolution. The dominant timescale of this evolution is ~300 fs, in good agreement with the wobbling times of other FA perovskites, demonstrating a correlation between the hydrogen bond dynamics and the rotational dynamics of the organic cation.

The complete rationalization for our experimental results is presented in Fig. 5. Figure 5a–c compares the kinetics of the experimentally obtained VIPER signal with three of the dynamic properties of the organic cation. Specifically, we compare the experimental VIPER kinetics (black dashed line) with the orientational anisotropy decay of the FA vibration in the $FAPbI_3$ perovskite[12] (Fig. 5a), the simulated dynamics of hydrogen bond breakage from AIMD (Fig. 5b) and the time-resolved cross-correlation between the Pb–Br and N–H•••X bond distance, which are reported elsewhere[41] (Fig. 5c). The PC-VIPER data broadly match both the experimental rotational dynamics observed via 2D-IR anisotropy decays and molecular dynamics results. In particular, the strongly negative cross-correlation (Fig. 5c) implies that a shortening of the N–H•••X bond distance is coordinated with an elongation of the Pb–Br bond distance. This suggests that hydrogen bonds between the organic cation and inorganic lattice induce a distortion of the latter. Thus, a picture emerges in which the N–C=N vibration of the cation affects N–H•••X hydrogen bonding, which, in turn, may alter the electronic states of the perovskite through distortion of the inorganic cage. The effect of the N–C=N vibration on hydrogen bonding may not be that surprising since the vibrational motion of the nitrogen atoms affects the hydrogen atoms involved with hydrogen bonding. Additionally, the relative softness and anharmonicity of HOIPs, combined with the rotationally mobile and anisotropic nature of the organic cation, opens up the possibility of dynamic fluctuations in the inorganic cage driven by the organic sublattice. The reverse process—in which inorganic lattice oscillations drive cation reorientation—has been invoked to explain differences in the rotational dynamics of the organic cation in a variety of HOIPs[12,43]. Some studies also suggest that hydrogen bonds between the organic cation and halide anions distort the $PbX_3$octahedra, indirectly influencing the material's optoelectronic behaviour[41]. We note, however, that our computational approach cannot account for the vibrational relaxation of the $\nu$(N–C=N) mode, which is a major loss pathway for our VIPER signals.

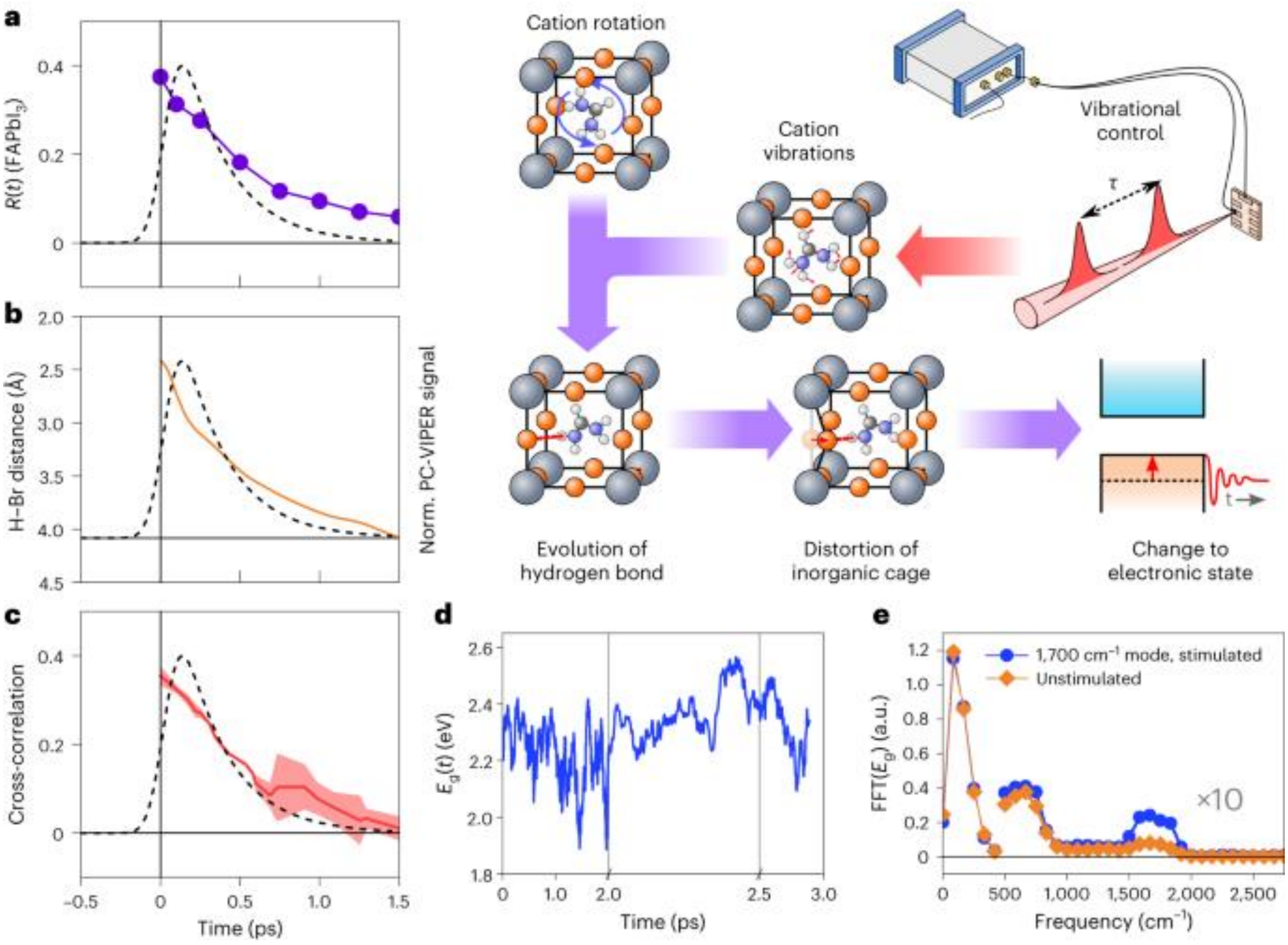

**Figure 5.** **a**–**c**, Comparison of PC-VIPER kinetics (dashed lines) to the previously experimentally determined rotational anisotropy of the FA ion in $FAPbI_3$ (**a**), the ab initio picosecond dynamics of the H–Br bond distance (**b**) and the time-resolved Pb–Br/H–Br cross-correlation function (mean ± standard deviation) obtained from the AIMD simulations performed elsewhere[41] (**c**). **d**,**e**, Analysis of the bandgap dynamics obtained from the AIMD simulations described in this work: bandgap of $FAPbBr_3$ for a representative trajectory following perturbation of the *ν*(N–C=N) mode (blue) (**d**); the power spectrum of the oscillatory component of bandgap fluctuations displayed in **d** (**e**), where the blue curve represents the average of the power spectra of the five trajectories in which the *ν*(N–C=N) mode is stimulated. The corresponding averaged power spectrum for the five unperturbed trajectories is given in orange.

The ab initio modelling at equilibrium is unable to evaluate the direct effect of the *ν*(N–C=N) dynamics of the organic cation on the electronic properties of the perovskite, as the amplitudes of the relevant mid-infrared (MIR; ~1,700 $cm^{-1}$) vibrational states are negligible at the simulated temperatures. We, therefore, modified the above approach by adjusting the atomic velocities along the normal modes corresponding to the *ν*(N–C=N) vibration of the FA ion, which we identify via phonon calculations. The adjustment amounts to three quanta of the *ν*(N–C=N) vibrational mode per supercell to simulate the IR-pulse excitation in the PC-VIPER and PL-VIPER experiments. In total, we ran ten individual trajectories—five in which the *ν*(N–C=N) mode was stimulated, and five in which no stimulation was introduced. The representative evolution of the perovskite bandgap for one such trajectory following the stimulation of the *ν*(N–C=N) modes is given in Fig. 5d, and shows both slow, large amplitude fluctuations and variation on faster timescales. Fourier analysis of the bandgap dynamics reveals the involvement of multiple modes at around 100, 700 and ~1,700 $cm^{-1}$ (Fig. 5e). We assign the modes in the <280 $cm^{-1}$ low-energy band to motions of the $PbBr_3$ octahedral cage, whereas the modes in the 300–1,000 $cm^{-1}$ region are commonly ascribed to hybrid modes of the organic cation rotations coupled to motions of the inorganic lattice[44]. The dominant mode is the oscillation at ~100 $cm^{-1}$, suggesting that the modulation of the bandgap is primarily driven by fluctuations of the inorganic cage. The mode at ~1,700 $cm^{-1}$ is probably the intramolecular *ν*(N–C=N) mode of FA, which matches well with the one we experimentally access by IR excitation in our VIPER experiments. Comparing the Fourier amplitude of the ~1,700 $cm^{-1}$ mode for trajectories with and without stimulation, we observe that the relative contribution of the *ν*(N–C=N) mode is strongly and selectively enhanced, highlighting its coupling to the bandgap of the perovskite. We note, however, that there are other important differences between the orientational selectivity of VIPER experiments versus our AIMD simulations arising from the polarization of the pulses that also must be considered (Supplementary Note 8).

Our results carry important implications for perovskite-based optoelectronic devices. Previous studies have emphasized the impact of the A-site cation on the electronic structure of HOIPs, inhibiting the cooldown of hot carriers[45]. Our findings elucidate the mechanistic picture behind this phenomenon. Moreover, we note that if high-frequency phonons can be used to stimulate sub-bandgap absorption, then the opposite pathway (electronic relaxation through intramolecular modes of the organic cation) may also be possible. Thus, the coupling of the high-frequency vibration of the cation to the electronic states in HOIPs introduces facile non-radiative pathways for carrier relaxation, which impacts the intrinsic non-radiative relaxation of photogenerated carriers. In HOIP solar cells, non-radiative losses have traditionally been the result of trap-mediated recombination, as well as losses at interfaces[46]. However, with the development of new defect passivation strategies, losses due to these extrinsic factors have reduced[47,48], making intrinsic non-radiative recombination dominant. Our findings suggest a natural route for controlling the coupling between the organic and inorganic sublattices of HOIPs (and therefore the intrinsic non-radiative loss) by controlling the strength of cation–halide hydrogen bond or/and the rotational mobility of the cation, in agreement with some recent findings[47]. Consequently, our study offers a new design guideline to perovskite device optimization that may push open-circuit-voltage ($V_{OC}$) losses down to their radiative limit, substantially improving their performance.

## Outlook

In conclusion, we have presented the first report of the ultrafast VC of an HOIP-based optoelectronic device. This experiment has provided access to the coupled electronic and structural dynamics of the $FAPbBr_3$ metal halide perovskite, both as a thin film and as a full optoelectronic device. Our spectroscopic techniques, namely, PC-VIPER and PL-VIPER, demonstrate that the $\nu$(N–C=N) stretching mode of the FA can modulate the (opto)electronic properties of $FAPbBr_3$. An analysis of the VIPER kinetics supports a picture in which coupling between the organic and inorganic sublattices results from a stochastic sticking and unsticking of the FA ion from the inorganic cage, mediated by cation–halide hydrogen bonding. Our results highlight the strongly dynamical nature of the coupling between the organic cation and inorganic cage and establish a direct link between this coupling and the optoelectronic performance of hybrid perovskites. The observed phenomena are potentially responsible for the reduction in the intrinsic non-radiative recombination of HOIPs, which impacts their performance as solar absorbers. Importantly, we suggest that the degree of intrinsic non-radiative recombinations can be controlled by optimizing the strength of hydrogen bonding and cation mobility. In the future, VC can be exploited to open new functionalities in ultrafast optical switches for time-division-multiplexed optical signals[49] as well as a powerful tool to study electron–phonon coupling, in both perovskites and a wider array of optoelectronic systems.

## Methods

### Fabrication of $FAPbBr_3$/$CsPbBr_3$ devices

The $FAPbBr_3$ and $CsPbBr_3$ devices employed in the PC-VIPER experiments were prepared in the configuration displayed in Fig. 1b, employing semitransparent Ag electrodes to enable the penetration of visible and IR pulses. The fabrication of perovskite devices was carried out in a dry, nitrogen-filled glovebox with a relative humidity of 0.5%. Initially, glass/indium tin oxide (PsiOTech, 15 Ω sq$^{-1}$) substrates were sonicated at a temperature of 50 °C, first in a soap and water solution and then in a mixture of deionized water, acetone and isopropanol. Thereafter, the substrates were subjected to a 10 min $O_2$-plasma treatment. A 1.5 mg ml$^{-1}$ solution of PTAA (Sigma-Aldrich) in toluene was prepared and spin coated onto the surface-treated glass/indium tin oxide substrates at 2,000 r.p.m. for 30 s; the resulting film was then annealed at 100 °C for 10 min. Thereafter, a 0.5 mg ml$^{-1}$ solution of poly(9,9-bis(3′-(*N*,*N*-dimethyl)-*N*-ethylammoinium-propyl-2,7-fluorene)-*alt*-2,7-(9,9-dioctylfluorene))dibromide (PFN-Br; Sigma-Aldrich) was spin coated onto the now-annealed PTAA film at 5,000 r.p.m. for 30 s.

To prepare the $FAPbBr_3$ active layer, FABr (GreatCell Solar) and $PbBr_2$ (TCI) were combined in a 1:1 molar ratio and dissolved in a 4:1 (v/v) *N*,*N*-dimethylformamide/dimethyl sulfoxide mixture, resulting in a 40 wt% solution of $FAPbBr_3$. This solution was subsequently spin coated onto the PTAA/PFN-Br layers, initially at 1,000 r.p.m. for 12 s, followed by 5,000 r.p.m. for 28 s. In the final 3 s of spin coating, 200 µl of anisol was dispensed onto the film as an antisolvent. The resulting film was annealed at 140 °C for 20 min.

In the case of $CsPbBr_3$ films, a previously reported two-step sequential deposition method was employed1. The substrates as well as a 1 M solution of $PbBr_2$ (TCI) in *N*,*N*-dimethylformamide were separately heated to 75 °C. The $PbBr_2$ solution was then spin coated onto the PTAA/PFN-Br film at 2,500 r.p.m. for 1 min. The resulting films were then dried for 30 min at 75 °C. To convert $PbBr_2$ into $CsPbBr_3$, the film was soaked in a 15 mg ml$^{-1}$ solution of CsBr (Sigma-Aldrich) in methanol for 10 min at 50 °C. The films were then dried and subsequently annealed at 250 °C for 10 min.

To apply the electron transport layers and Ag electrodes, an identical procedure was employed for both $FAPbBr_3$ and $CsPbBr_3$ films. A 20 mg ml$^{-1}$ solution of [6,6]-phenyl-$C_{61}$-butyric acid methyl ester (Solenne) in chlorobenzene was prepared and then passed through a 0.25 µm polytetrafluoroethylene filter to remove the residual solid particulate. The solution was then dynamically cast onto the active layers at 2,000 r.p.m. for 30 s. The film was then annealed at 100 °C for 10 min. A bathocuprene (Sigma-Aldrich) hole-blocking layer was then spin coated onto the [6,6]-phenyl-$C_{61}$-butyric acid methyl ester film from a 0.5 mg ml$^{-1}$bathocuprene/isopropanol solution at a speed of 4,000 r.p.m. for 30 s. After drying, silver electrodes were evaporated onto the film at a rate of 0.1 Å s$^{-1}$ to a thickness of 18 Å.

**Fabrication of nanocrystalline $FAPbBr_3$ films**

To produce the oleylammonium bromide (OLAmBr) precursor, 12.5 ml oleylammonium was mixed with 100 ml EtOh in a flask and cooled in an ice/water bath. Thereafter, 8.56 ml HBr (48% aqueous solution) was added. The reaction mixture was left to react overnight under a nitrogen flow. The resulting solution was dried in a rotary evaporator and the remaining precipitate was washed multiple times with diethyl ether. The resultant white OLAmBr powder was dried under a vacuum at room temperature for several hours and then stored in a glovebox.

To synthesis the $FAPbBr_3$ NCs, 76 mg lead(II) acetate trihydrate (0.20 mmol) and 78 mg formamidinium acetate (0.75 mmol) were dispersed in a mixture of 8 ml ODE and 2 ml OAc, heated to 50 °C and then dried under a vacuum for 30 min. Thereafter, the reaction mixture was heated to 130 °C and a separately prepared solution of 266 mg OLAmBr (0.8 mmol) in 2 ml anhydrous toluene was injected into the reaction flask. To prepare the OLAmBr/toluene solution, the mixture was heated to 40–50 °C and held at this temperature for approximately 30 s, after which the reaction mixture was cooled in an ice/water bath.

To obtain purified NCs from the resulting crude solution, 16 ml of methyl acetate was added and the mixture was centrifuged at 12,100 r.p.m. for 5 min. The resulting supernatant was discarded and the remaining precipitate was redissolved in toluene (5 ml). This solution was centrifuged a second time at 3,000 r.p.m. for 2 min. In this case, the supernatant was retained for the didodecylammonium bromide treatment, and any precipitated NCs were discarded.

To treat the purified $FAPbBr_3$ NCs with didodecylammonium bromide, 5 ml toluene was added to the prepared $FAPbBr_3$ colloidal solution as described above. Thereafter, 0.10 ml of OAc and 0.54 ml of didodecylammonium bromide (0.05 M in toluene) were added to 10.00 ml colloidal solution of $FAPbBr_3$ NCs; the resulting mixture was stirred for 1 h followed by precipitation with 16.00 ml ethyl acetate. The resulting suspension was centrifuged at 12,100 r.p.m. for 3 min. The precipitate was redispersed in 2.5 ml toluene and this solution was additionally filtered through a 0.45 μm polytetrafluoroethylene filter.

To deposit the $FAPbBr_3$ NCs, $CaF_2$ substrates (12 mm diameter, 1 mm thickness; EKSMA Optics) were treated with $O_2$ plasma to improve wetting. The concentrated NC inks were drop cast from the solution onto the cleaned substrates in a $N_2$atmosphere. The thickness of the thin films were determined to be ~300 nm by UV–visible ellipsometry.

**PC-VIPER and PL-VIPER experiments**

The primary light source for the experiments was the output from a Ti:sapphire regenerative amplifier (Astrella, Coherent), which outputs 800 nm pulses at a repetition rate of 4 kHz, with a nominal temporal width of 35 fs. The output of the amplifier is divided and passed onto a pair of optical parametric amplifiers (OPAs; TOPAS Prime, Coherent), which produce near-infrared (NIR) signal and idler pulses in the 1,140–2,671 nm range. To prepare the MIR pulses for the pre-excitation step, the output from one OPA was passed to a custom-built difference-frequency generation rig, which uses an $AgGaS_2$ crystal to achieve MIR pulses tunable in the 2,857–6,666 nm (1,500–3,500 $cm^{-1}$) range. In the experiments described in this Article, the difference-frequency generation was tuned to output pulses with a centre frequency of ~1,720 $cm^{-1}$, and a full-width at half-maximum of approximately 200 $cm^{-1}$; Supplementary Figs. 1 and 2 show a detailed spectrum of the pulse. The residual NIR light from the OPA was removed by means of a long-pass dielectric filter with a cut-on wavelength of 3,000 nm (LP-3000, Spectrogon).

To produce the visible pulses for the secondary electronic excitation, we made use of second-harmonic generation, using the 1,150 nm output from the second OPA and a beta barium borate crystal (BBO-0110-06H, EKSMA Optics UAB). The residual NIR light and any other unwanted light (for example, from OPA superfluorescence) was removed by means of a series of dielectric filters (Thorlabs) to yield pulses centred at ~575 nm (Fig. 2). To avoid the addition of optics that could degrade the temporal resolution of the VIPER system, the polarization of both IR pre-excitation and visible electronic excitation pulses were not altered, resulting in the pulses being parallelly polarized with respect to each another. Cross-polarization of the pulses would not be expected to provide additional structural information, as the direction(s) of the electronic and vibrational transition dipoles are both random and isotropic (both with respect to the laboratory frame of reference and with respect to one another) within the length scales probed by the visible and NIR pulses ($10^2$–$10^3$ μm).

After the initial preparation, the MIR and visible pulses were passed to the experimental setup shown in Fig. 1. The system was sealed and purged with dry air to reduce the relative humidity to below 1%. To achieve temporal resolution, the visible pulse was passed to a motorized delay stage (DDS220/M, Thorlabs), with a total travel range of ~220 nm. To achieve spectral resolution, the MIR pulse is passed to a custom-built interferometer. To accurately measure the position of the interferometer delay line, a 635 nm reference laser was also passed through the interferometer, enabling submicrometre position determination. To prevent picket fencing of the reference interferogram, the interferogram was deliberately oversampled; the approximate spacing between the adjacent sample points was 10 nm and the dwell time at each interferogram point was approximately 10 ms (40 laser shots per sample point).

After the interferometer, the visible and IR beams were combined at a germanium dielectric filter (WG91050-C9, Thorlabs) and subsequently focused onto the sample. The MIR pulses were modulated at 2 kHz using an optical chopper (MC2000B-EC, Thorlabs) equipped with a 30-slot blade (MC1F30, Thorlabs). The average power of the IR and visible pulses at the sample position was approximately 300 and 50 μW, respectively. For the visible pulses, this corresponded to a pulse energy of 12.5 nJ, whereas for the MIR pulses, the power was evenly divided across the static and dynamic pulse image trains, resulting in pulse energies of 75 nJ per pulse. The focal-spot diameters of both visible and NIR beams at the sample were approximately 0.5 mm. The sample device was mounted on a closed, custom-designed sample holder, equipped with a $CaF_2$ window under a positive nitrogen pressure for the entire experiment. To detect the photocurrent, the device and optical chopper were connected to a lock-in amplifier (MFLI, Zurich Instruments), operating at a time constant of 5 ms and a sensitivity of 1 μA. The interferometer, delay lines and detection system were controlled using custom control software written with LabVIEW 2019.

The same setup was used for the PC-VIPER and PL-VIPER experiments; however, the detection system was slightly altered (Supplementary Note 3).

The resulting datasets were analysed using OriginPro (2021b) and MATLAB (R2021b). The resulting graphs were plotted using OriginPro (2021b)

**AIMD simulations**

We perform Born–Oppenheimer AIMD with the QUANTUM ESPRESSO code[50,51]. A $2 \times 2 \times 2$ supercell containing eight formula units of $FAPbBr_3$ is chosen. The lattice constants were fully relaxed for a supercell volume of 2,025 $Å^3$ and then fixed for the remainder of the calculations. We use the Perdew–Burke–Ernzerhof[52] functional and the Tkatchenko–Scheffler scheme[53] for dispersive interactions. A $2 \times 2 \times 2$ Monkhorst–Pack *k*-point grid is used[54], and the planewave cutoff energy is 50 Ry. The density-functional-theory-calculated total energy is converged to $10^{-7}$ Ry per cell. Spin–orbit coupling is not considered, as it does not affect the structural properties of lead halide perovskites. Equilibrium AIMD was performed in the canonical (*NVT*) ensemble at 300 K with velocities integrated using the Verlet algorithm with a time step of 1.5 fs. Temperature was controlled using a Berendsen thermostat55 with a time constant of 150 fs. The trajectory was equilibrated for 2 ps and further 50 ps of dynamics were then used for analysis. Although velocity-rescaling thermostats such as the Berendsen thermostat can sometimes incorrectly partition energy between the modes of different frequencies, they do not introduce artefacts in dynamical processes as velocity randomization thermostats are known to do56. Although thermostating introduces an additional characteristic time into the simulation, this timescale does not appear in any of the correlation properties calculated; therefore, we are confident that the computed dynamics are a real property of the system. The resulting datasets were analysed using Python 3.10 and plotted using Matplotlib 3.5.

**Acknowledgements**

We thank M. Pchenitchnikov, T. L. C. Jansen, D. Egger, D. Klug and D. Cahen for useful discussions. We thank O. D. Parashchuk for helping in conducting the Raman measurements. N.P.G. thanks T. Shibu for help with the preliminary VIPER measurements. K.P.G. and Y.V. thank A. D. Taylor for help with developing the fabrication procedures of the FAPbBr3 devices. D.R.M. acknowledges funding by the Imperial College London President's PhD Scholarships. N.M. and A.A.B. acknowledge support from the European Commission through the Marie Skłodowska-Curie Actions under Horizon 2020 (Project PeroVIB, 101018002). A.A.B. is a Royal Society University Research Fellow. A.A.B. acknowledges support from Leverhulme Trust via Philip Leverhulme Prize award. This project received funding from the European Research Council (ERC) under the European Union's Horizon 2020 research and innovation programme (grant agreements 639750/VIBCONTROL and 714067/ENERGYMAPS). Y.V. also thanks the Deutsche Forschungsgemeinschaft (DFG) for funding the 'PERFECT PVs' project (grant no. 424216076) in the framework of SPP 2196. M.I.B. acknowledges financial support from the Swiss National Science Foundation (grant no. 200021_192308, Project Q-Light). T.T. acknowledges support from JSPS KAKENHI (grant nos. 18H05265 and 23H00292). The theoretical work (Z.D., A.M.S. and A.M.R.) was supported by the National Science Foundation, Science and Technology Center Program (IMOD), under grant no. DMR-2019444. Computational support was provided by the National Energy Research Scientific Computing Center (NERSC), a US Department of Energy, Office of Science User Facility, located at Lawrence Berkeley National Laboratory, operated under contract no. DE-AC02-05CH11231.